\documentclass[twocolumn,showpacs]{revtex4}

\usepackage{graphicx}
\usepackage{dcolumn}
\usepackage{bm}

\begin{document}

\preprint{LC9183}

\title{Mott Gap Excitations and Resonant Inelastic X-Ray Scattering in Doped Cuprates}

\author{Kenji \surname{Tsutsui}}
\author{Takami \surname{Tohyama}}
\author{Sadamichi \surname{Maekawa}}
\affiliation{%
Institute for Materials Research,
Tohoku University, Sendai 980-8577, Japan
}%

\date{13 March 2003}
\begin{abstract}
Predictions are made for the momentum- and carrier-dependent degradation of the Mott gap upon doping in high-$T_\mathrm{c}$ cuprates as would be observed in Cu $K$-edge resonant inelastic x-ray scattering (RIXS).
The two-dimensional Hubbard model with second- and third-nearest-neighbor hopping terms has been studied by numerical exact diagonalization.
Special emphasis is placed on the particle-hole asymmetry of the Mott gap excitations.
We argue that the Mott gap excitations observed by RIXS are significantly influenced by the interaction between charge carriers and antiferromagnetic correlations.
\end{abstract}

\pacs{74.25.Jb, 71.10.Fd, 78.70.Ck}
\maketitle

Asymmetry in the electronic properties between electron- and hole-doped high-$T_\mathrm{c}$ cuprates is one of the key issues relevant to an understanding of the mechanism of high-$T_\mathrm{c}$ superconductivity.
In the cuprates, the antiferromagnetic (AF) correlations are stronger in the electron-doped than hole-doped systems \cite{takagi}: In the electron-doped Nd$_{2-x}$Ce$_x$CuO$_4$ (NCCO), the antiferromagnetism continues up to $x=0.15$, whereas it disappears for a much smaller $x$ ($\sim0.02$) in hole-doped La$_{2-x}$Sr$_x$CuO$_4$ (LSCO).
The kind of carriers is reflected not only in the magnetic properties but also in the single-particle excitations.
Angle-resolved photoemission spectroscopy (ARPES) experiments revealed that while the minimum energy excitation appears at around $(\pi/2,\pi/2)$ for hole doping, it is seen near $(\pi,0)$ for electron doping \cite{damascelli}.
On the other hand, the charge-transfer excitations from the occupied Zhang-Rice singlet band (ZRB) \cite{zhang} composed of Cu 3$d_{x^2-y^2}$ and O 2$p_\sigma$ orbitals to the unoccupied upper Hubbard band (UHB), i.e. Mott gap excitation, seen in the optical conductivity, shows similar behavior for electron and hole dopings \cite{uchida,arima}.
In the optical process, the excitations occur without the momentum transfer to the electron system. 
Resonant inelastic X-ray scattering (RIXS) is now gaining importance as a powerful technique for the investigation of the momentum-dependent excitations with the energies of the order of the Mott gap energy \cite{hill,hasan1d,tsutsui1d,hasan,kim,tsutsui}.
In particular, the Cu $K$-edge RIXS measurements on the parent materials of high-$T_\mathrm{c}$ cuprates \cite{hasan,kim} revealed the characteristic momentum dependence of the Mott gap excitation \cite{tsutsui}.
It is, thus, interesting to clarify whether a doping asymmetry also exists in this high-energy region and how the nature of the Mott gap excitations in doped cuprates looks in RIXS.

In this Letter, we demonstrate theoretically the difference of the Cu $K$-edge RIXS between hole- and electron-doped cuprates.
The single-band Hubbard model with long-range hoppings is adopted to describe the ZRB and UHB by mapping ZRB onto the lower Hubbard band (LHB) in the model.
Then, the Cu $1s$ and $4p$ orbitals are incorporated with the model to include the $1s$-core hole and excited $4p$ electron into the intermediate state of the RIXS process.
The RIXS spectra are calculated by using the numerically exact diagonalization technique.
We find that the excitation spectrum from the LHB to the UHB becomes broad and less momentum dependent upon hole doping.
This is in contrast to the electron-doped case, where the momentum dependence of the spectrum of undoped system remains, except that along the $\langle 1,0 \rangle$ direction.
The difference in the spectra between hole- and electron-doped systems follows the carrier-dependence of short-range AF spin correlation. 
We also find that the momentum dependence along the $\langle 1,0\rangle$ direction in electron doping comes from the fact that the doped electrons occupy the states at around $\mathbf{k}=(\pi,0)$ in the upper Hubbard band.

By mapping the ZRB onto the LHB, which is equivalent to the elimination of O $2p$ orbitals \cite{tsutsui}, the Hubbard Hamiltonian with second- and third-nearest-neighbor hoppings is written as,
\begin{eqnarray}
\label{ham3d}
H_{3d} &=& -t\sum_{\langle \mathbf{i,j} \rangle_{\text{1st}}, \sigma}
          d_{\mathbf{i},\sigma}^\dag d_{\mathbf{j},\sigma}
        -t'\sum_{\langle \mathbf{i,j} \rangle_{\text{2nd}}, \sigma}
          d_{\mathbf{i},\sigma}^\dag d_{\mathbf{j},\sigma} 
  \nonumber\\&&
      -t''\sum_{\langle \mathbf{i,j} \rangle_{\text{3rd}}, \sigma}
          d_{\mathbf{i},\sigma}^\dag d_{\mathbf{j},\sigma} + \text{h.c.}
      +U\sum_\mathbf{i}
          n^d_{\mathbf{i},\uparrow}n^d_{\mathbf{i},\downarrow},
\end{eqnarray}
where $d_{\mathbf{i},\sigma}^\dag$ is the creation operator of $3d$ electron with spin $\sigma$ at site $\mathbf{i}$, $n_{\mathbf{i},\sigma}^d=d_{\mathbf{i},\sigma}^\dag d_{\mathbf{i},\sigma}$, the summations $\langle \mathbf{i,j} \rangle_{\text{1st}}$, $\langle \mathbf{i,j} \rangle_{\text{2nd}}$, and $\langle \mathbf{i,j} \rangle_{\text{3rd}}$ run over first, second, and third nearest-neighbor pairs, respectively, and the rest of the notation is standard.
The on-site Coulomb energy $U$ corresponds to the charge transfer energy of cuprates.

\begin{figure}[t]
\includegraphics[width=7.5cm]{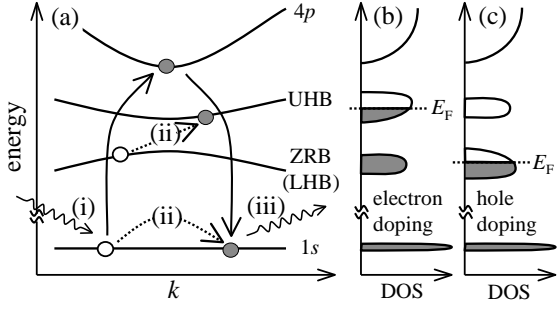}
\vspace{-10pt}
\caption{\label{fig1}
Schematic picture of Cu $K$-edge RIXS process (a) and densities of states for electron- (b) and hole-doped (c) cases.
The dipole transitions of an electron between $1s$ and $4p$ orbitals occur with (i) absorption and (iii) emission of a photon.
The $3d$ electrons are excited (ii) in the intermediate sate.
The Fermi energies [denoted by dotted lines in (b) and (c)] are in UHB and ZRB (LHB) for electron- and hole-doped cases, respectively.
}
\end{figure}

The schematic process of the RIXS for Cu $K$-edge is shown in Fig. \ref{fig1} (a).
An absorption of an incident photon with energy $\omega_\mathrm{i}$, momentum $\mathbf{K}_\mathrm{i}$, and polarization $\bm{\epsilon}_\mathrm{i}$ brings about the dipole transition of an electron from Cu $1s$ to $4p$ orbital [process (i)].
In the intermediate states, $3d$ electrons interact with a $1s$-core hole and a photoexcited $4p$ electron via the Coulomb interactions so that the excitations in the $3d$ electron system are evolved [process (ii)].
The Fermi energies are located in the UHB and ZRB for electron- and hole-doped cases, respectively.
As shown in Figs. \ref{fig1}(b) and \ref{fig1}(c), the excitations occur within the band as well as between the bands.
The latter corresponds to the Mott gap excitation.
The $4p$ electron in the intermediate state goes back to the $1s$ orbital again and a photon with energy $\omega_\mathrm{f}$, momentum $\mathbf{K}_\mathrm{f}$, and polarization $\bm{\epsilon}_\mathrm{f}$ is emitted [process (iii)].
The differences of energies and momenta between incident and emitted photons are transferred to the $3d$ electron system.

In the intermediate states, there are a $1s$-core hole and a $4p$ electron, with which $3d$ electrons interact.
Since the $1s$-core hole is localized in a small radius of the Cu $1s$ orbital, the attractive interaction between the $1s$-core hole and $3d$ electrons is strong.
The interaction is written as,
$
H_{1s\text{-}3d}=-U_\mathrm{c}\sum_{\mathbf{i},\sigma,\sigma'}
n_{\mathbf{i},\sigma}^d n_{\mathbf{i},\sigma'}^s,
$
where $n_{\mathbf{i},\sigma}^s$ is the number operator of $1s$-core hole with spin $\sigma$ at site $\mathbf{i}$, and $U_\mathrm{c}$ is taken to be positive.
On the contrary, the Coulomb interactions related to the $4p$ electron are neglected since the $4p$ electron is delocalized \cite{tsutsui}.
Furthermore, we assume that the photoexcited $4p$ electron enters into the bottom of the $4p$ band with momentum $\mathbf{k}_0$.
Under these assumptions, the RIXS spectrum is expressed as,
\begin{eqnarray}\label{rixs}
I(\Delta \mathbf{K},\Delta\omega)&=&\sum_\alpha\left|\langle\alpha|
\sum_\sigma s_{\mathbf{k}_0-\mathbf{K}_\mathrm{f},\sigma} p_{\mathbf{k}_0,\sigma}
 \right.\nonumber\\&&\times\left. 
\frac{1}{H-E_0-\omega_\mathrm{i}-i\Gamma} p_{\mathbf{k}_0,\sigma}^\dag s_{\mathbf{k}_0-\mathbf{K}_\mathrm{i},\sigma}^\dag |0\rangle\right|^2
\nonumber\\&&\times
\delta(\Delta\omega-E_\alpha+E_0),
\end{eqnarray}
where $H=H_{3d}+H_{1s\text{-}3d}+H_{1s,4p}$, $H_{1s,4p}$ being composed of the energy separation $\varepsilon_{1s\text{-}4p}$ between the $1s$ level and the bottom of the $4p$ band, $\Delta\mathbf{K}=\mathbf{K}_\mathrm{i}-\mathbf{K}_\mathrm{f}$, $\Delta\omega=\omega_\mathrm{i}-\omega_\mathrm{f}$, $s_{\mathbf{k},\sigma}^\dag$ ($p_{\mathbf{k},\sigma}^\dag$) is the creation operator of the $1s$-core hole ($4p$ electron) with momentum $\mathbf{k}$ and spin $\sigma$, $|0\rangle$ is the ground state with energy $E_0$, $|\alpha\rangle$ is the final state of the RIXS process with energy $E_\alpha$, and $\Gamma$ is the inverse of the relaxation time in the intermediate state.
The RIXS spectrum of Eq. (\ref{rixs}) is calculated on a $4\times4$-site cluster with periodic boundary conditions by using a modified version of the conjugate-gradient method together with the Lancz\"os technique.

The values of the parameters are as follows:
$t'/t=-0.34$, $t''/t=0.23$, $U/t=10$, $U_\mathrm{c}/t=15$, and $\Gamma/t=1$ with $t=0.35$ eV that are estimated from the analyses of ARPES data \cite{kimpes} and are the same as those used in Ref.~\cite{tsutsui}.

\begin{figure}[t]
\includegraphics[width=8cm]{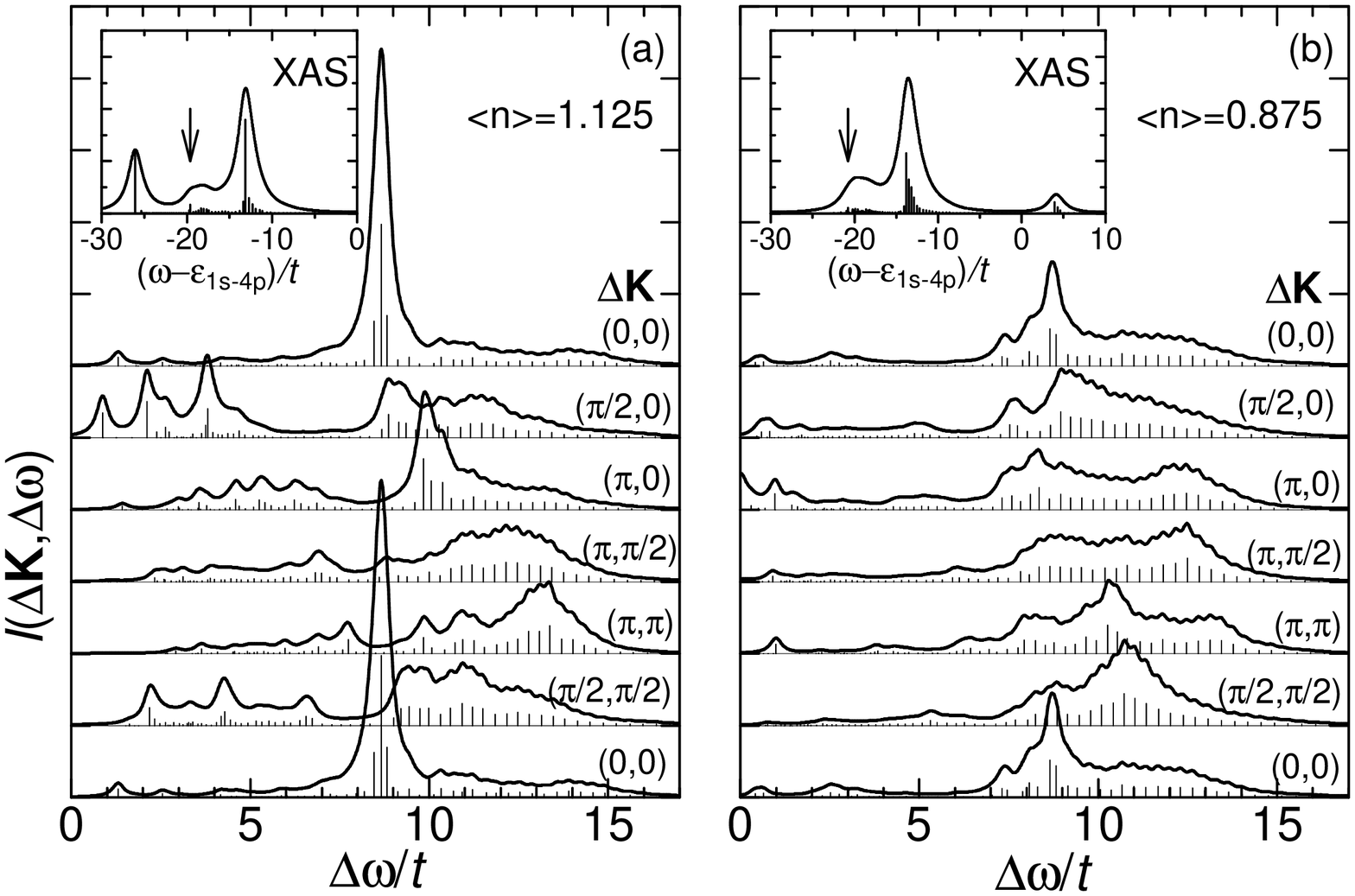}
\vspace{-15pt}
\caption{\label{figdope}
Resonant inelastic x-ray scattering spectra (RIXS) for Cu $K$-edge of doped Hubbard model with long-range hoppings.
(a) The electron doped case ($\langle n \rangle=1.125$). (b) The hole doped case ($\langle n \rangle=0.875$).
The spectra of the elastic scattering process at $\Delta \mathbf{K} = (0,0)$ are not shown.
The parameters are $U/t=10$, $U_\mathrm{c}/t=15$, $\Gamma/t=1$, $t'/t=-0.34$, and $t''/t=0.23$.
The $\delta$-functions (the vertical thin solid lines) are convoluted with Lorentzian broadening of $0.2t$.
Insets are the Cu 1$s$ absorption spectra with $\Gamma_{\text{XAS}}/t=\Gamma/t=1.0$, and the incident photon energies $\omega_\mathrm{i}$'s for RIXS are set to the values denoted by the arrows.
}
\end{figure}

Insets of Figs. \ref{figdope}(a) and \ref{figdope}(b) are the Cu $1s$ x-ray absorption spectrum (XAS) for electron- and hole-doped cases, respectively \cite{state}.
The spectrum is given by
\begin{eqnarray}
D(\omega)&=&\frac{1}{\pi}\text{Im}\langle 0 | s_{\mathbf{k}_0-\mathbf{K},\sigma} p_{\mathbf{k}_0,\sigma} \frac{1}{H-E_0-\omega-\mathrm{i}\Gamma_{\text{XAS}}} \nonumber\\&&\times
p_{\mathbf{k}_0,\sigma}^\dag s_{\mathbf{k}_0-\mathbf{K},\sigma}^\dag |0\rangle,
\end{eqnarray}
where $H$ is the same as that in Eq.~(\ref{rixs}).
In the inset of Fig.~\ref{figdope} (a), there appear three peaks, i.e., peaks at around $\omega-\varepsilon_{1s\text{-}4p}=-26t$, $-20t$, and $-13t$.
The peaks at around $-20t$ and $-13t$ are also seen in the undoped case \cite{tsutsui}.
The peak at around $-20t$ corresponds to a final state where the core hole is screened and thus the core-hole site is doubly occupied by $3d$ electrons ($U-2U_\mathrm{c}=-20t$).
This final state promotes the excitations from LHB to UHB in the RIXS.
The peak at around $-13t$ represents unscreened core-hole state and mainly contains the configuration that a core hole is created at a singly occupied site ($-U_\mathrm{c}=-15t$).
The peak at around $-26t$, which appears upon electron doping, corresponds to a final state where the core-hole is created at a doubly occupied site induced by electron doping ($-2U_\mathrm{c}=-30t$).
In the hole-doped case in Fig.~\ref{figdope}(b), a peak appears at around $4t$, in addition to the peaks at $-20t$ and $-13t$.
The peak at around $4t$ mainly contains the configuration that the core-hole is created at
an empty site induced by hole doping.
Because we are interested in the excitations from LHB to UHB in the RIXS spectra, we set $\omega_\mathrm{i}$'s to the energy of the peak at around $-20t$ as denoted by the arrows in the insets of Figs.~\ref{figdope}(a) and \ref{figdope}(b).

Figure~\ref{figdope}(a) shows the RIXS spectra in the electron-doped case ($\langle n \rangle=1.125$, where $\langle n \rangle$ is the electron concentration per site).
The spectra below $\Delta \omega \sim 8t$ are associated with the excitations within the UHB, and those above $8t$ are due to the excitations from LHB to UHB, i.e., Mott gap excitations.
We find that, in the energy region above $8t$, the spectra strongly depend on momentum, showing a feature that the spectral weight shifts to higher energy region with increasing $\left|\Delta\mathbf{K}\right|$.
The feature is similar to that of undoped case \cite{tsutsui}.
We note that, when the incident energy $\omega_\mathrm{i}$ is set to $\sim -26t$ which is the energy of the lowest-energy peak in XAS, the RIXS spectra appear only in the lower energy region below $\Delta\omega\sim 8t$ (not shown).
This is because the excitation from a singly occupied state to a doubly occupied state is reduced in the intermediate state of the RIXS process.
Therefore, by tuning $\omega_\mathrm{i}$ to a peak energy of $U-2U_\mathrm{c}\sim -20t$, we can enhance strongly the RIXS spectra associated with the Mott gap excitations.
Since there is no such an enhancement effect on the Mott gap excitation in the optical conductivity \cite{uchida,arima} as well as electron-energy loss spectroscopy, we can say that by utilizing the incident energy dependence, the RIXS is the unique technique to investigate the Mott gap excitation even in doped materials.

Figure~\ref{figdope}(b) shows the RIXS spectra in the hole-doped case ($\langle n \rangle=0.875$).
The spectra above the energy $\sim 6t$ are associated with the Mott gap excitations.
Compared with the electron-doped case, the spectra show broad features:
For example, at $\Delta \mathbf{K}=(\pi,\pi)$, the spectrum extends from $\Delta\omega\sim 7t$ to $\sim 14t$, and the energy position of the maximum spectral weight ($\sim 10t$) is lower than that of electron-doped case ($\sim 13t$).
The spectra at other $\Delta \mathbf{K}$'s are also extended to wide energy region similar to that at $(\pi,\pi)$, and the energy distributions of spectral weights seem to be rather independent of momentum.

\begin{figure}[t]
\includegraphics[width=7cm]{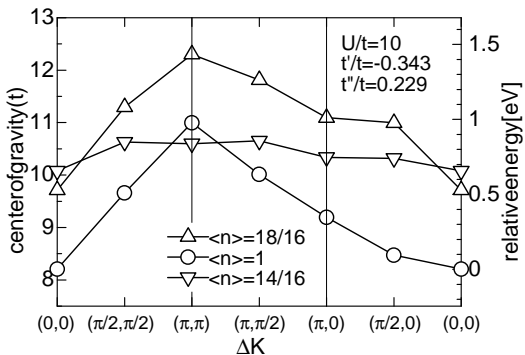}
\vspace{-15pt}
\caption{\label{figgra1}
Momentum dependence of the center of gravity of RIXS spectra associated with the excitation from LHB to UHB.
The spectral weight is adopted in energy regions above $4t$, $8t$, and $6t$ for undoped ($\langle n\rangle=1$, denoted by circles), electron- (1.125, upward triangles), and hole-doped (0.875, downward triangles) cases, respectively.
The right axis denotes the relative energy from the energy at $\Delta \mathbf{K}=(0,0)$ with $t=0.35$ eV.
}
\end{figure}

In order to compare the energy positions of the spectra, the momentum dependence of the center of gravity of the RIXS spectra associated with the Mott gap excitation is plotted in Fig.~\ref{figgra1}, where the spectral weight is adopted from the energy regions above $4t$, $8t$, and $6t$ for undoped ($\langle n\rangle=1$, denoted by circles) \cite{tsutsui}, electron- (1.125, upward triangles), and hole-doped (0.875, downward triangles) cases, respectively.
We find that the energy positions in the doped cases are shifted to the high-energy side compared with the undoped case.
This is because, upon hole (electron) doping, the Fermi energy shifts to LHB (UHB) and the energy difference between states of the occupied LHB and unoccupied UHB becomes large.
In the undoped case, the center of gravity shifts to higher energy with increasing $\left|\Delta\mathbf{K}\right|$.
It has been shown that this momentum dependence has good correspondence with the experimental data \cite{hasan}.
We find that the momentum dependence in the undoped case remains by electron doping, except along the $\langle 1,0 \rangle$ direction where the energy difference between $(\pi/2,0)$ and $(0,0)$ is larger than that in the undoped case.
The origin of the similarity of the momentum dependence will be discussed below.
Here let us discuss the physics behind the behavior along the $\langle 1,0 \rangle$ direction.
In the undoped case, the RIXS spectra have a characteristic feature along the $\langle 1,0 \rangle$ direction where the edge of the RIXS spectrum at $\Delta \mathbf{K}=(\pi/2,0)$ is rather lower in energy than that at $(0,0)$ \cite{tsutsui}:
The edge of the RIXS at $\Delta \mathbf{K}=(\pi/2,0)$ comes from the excitation from the occupied $\mathbf{k}=(\pi/2,0)$ state to the unoccupied $\mathbf{k}=(\pi,0)$ one which is the lowest in energy in UHB \cite{tsutsui}.
This edge feature restrains the center of gravity at $(\pi/2,0)$ from shifting to the higher energy in the undoped case.
Since the $(\pi,0)$ state of UHB is occupied by electrons upon electron doping, the center of gravity at $\Delta \mathbf{K}=(\pi/2,0)$ in the RIXS shifts to the higher energy.

\begin{figure}[t]
\includegraphics[width=6.7cm]{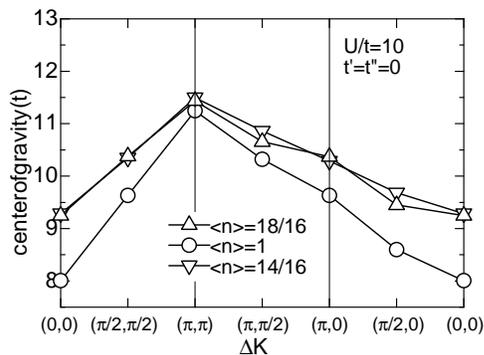}
\vspace{-15pt}
\caption{\label{figgra2}
The same as Fig.~\ref{figgra1} but for $t'=t"=0$.
The spectral weight is adopted in energy regions above $6t$, $7t$, and $7t$ for undoped, electron-, and hole-doped cases, respectively.
}
\end{figure}

In contrast, the momentum dependence becomes weaker by hole doping, as seen in Fig.~\ref{figgra1} \cite{ten}.
Let us make clear that the difference between electron and hole dopings comes from the lack of particle-hole symmetry in cuprates but not from the RIXS process itself in which the background configurations around the core-hole screened site are different between electron and hole dopings.
On Fig.~\ref{figgra2} is plotted the center of gravity of RIXS spectrum on the Hubbard model with only the nearest-neighbor hopping, which has the particle-hole symmetry.
To obtain the RIXS spectra, the incident photon energies $\omega_\mathrm{i}$'s are set to the peak energy of $U-2U_\mathrm{c}$ in XAS spectra for both electron- and hole-doped cases.
We can obtain almost the same momentum dependence in both electron and hole dopings accompanied by the same energy shift from the undoped case, although the configurations in the intermediate states are different.
Therefore, the spectra of the Mott gap excitations have almost the same features between electron and hole dopings when the model has the particle-hole symmetry.

One possible origin of the fact that the momentum-dependent spectra remain upon electron doping, whereas the broad feature appears in the hole-doped case, comes from the doping effect on the AF correlations.
In the undoped case, it has been discussed that the AF correlation plays a crucial role in the RIXS excitation \cite{tsutsui,tsutsui2}:
The matrix elements of the excitation from the top of the LHB [$\mathbf{k}=(\pi/2,\pi/2)$] to the bottom of the UHB [$\mathbf{k}=(\pi,0)$] are almost zero due to the coherence factor arising from the AF order.
This fact affects strongly the momentum dependence of RIXS in the undoped case.
Upon doping, the AF correlation changes.
However, it strongly depends on carriers in the presence of the long-range hoppings \cite{tohyamasq,godding,tohyamael}:  With the parameters used in Fig.~\ref{figdope}, the short-range AF correlation is kept in the electron-doped case, whereas the correlation is strongly suppressed in the hole-doped case.
Thus, we can suppose that the RIXS spectra in the electron-doped system are similar to the undoped case, while the spectra in the hole-doped system are different.
In fact, as discussed above, in the electron-doped case, the calculated RIXS spectra show the momentum dependence similar to that of the undoped case, but remarkably different in the hole-doped case.
Therefore, it is natural to consider that the Mott gap excitation is significantly influenced by the magnitude of AF correlations even in doped systems.

Finally, we discuss the material dependence of RIXS.
The parameters of the second- and third-nearest-neighbor hoppings, $t'$ and $t''$, play a role in the material dependence of the electronic properties.
The hopping parameters $t'$ and $t''$ in LSCO are smaller than those in Bi$_2$Sr$_2$CaCu$_2$O$_8$ (BSCCO), Ca$_2$CuO$_2$Cl$_2$ (CCOC), and NCCO \cite{tohyama}.
Therefore the features of the RIXS spectra on BSCCO, CCOC and NCCO are expected to follow Fig.~\ref{figgra1} \cite{hasanprivate}, while for doped LSCO, the feature may be rather similar to that in Fig.~\ref{figgra2} \cite{kimprivate}.

In summary, we have demonstrated theoretically the difference of the Cu $K$-edge RIXS between hole- and electron-doped cuprates by using the numerically exact diagonalization technique on small clusters of the single-band Hubbard model with long-range hoppings.
We have found that, upon electron doping, the RIXS spectra along the $\langle 1, 0 \rangle$ direction shift to the higher energy side than those in undoped case.
In contrast to electron-doped case, the spectra for the excitations from the lower to upper Hubbard bands show the broad feature and less momentum dependence in the hole-doped case.
The difference in the spectra between hole- and electron-doped systems follows the carrier dependence of short-range AF spin correlation. 
The RIXS experiments in a variety of doped cuprates are desired \cite{kimprivate,hasanprivate}.

The authors would like to thank M. Z. Hasan, J. P. Hill, and Y. J. Kim for informing their experimental data prior to publication and for valuable discussions.
The numerical calculations in Fig.~\ref{figdope} are assisted by H. Kondo.
This work was supported by Priority-Areas Grants from the Ministry of Education, Science, Culture and Sport of Japan and CREST.
Computations were carried out in ISSP, University of Tokyo; IMR, Tohoku University; and Tohoku University.


\begin{thebibliography}{}
\bibitem{state}
For the calculation of the spectra, we adopt the initial states with $d$-wave symmetry in both electron- and hole-doped cases.
\bibitem{ten}
The same trend of the carrier dependence on RIXS is obtained in 10\%-doped systems of a two-dimensional 10-site cluster.
\end{thebibliography}
\end{document}